\documentclass[utf8]{FrontiersinHarvard} 

\usepackage[onehalfspacing]{setspace}
\usepackage{cite}
\usepackage{amsmath,amssymb,amsfonts}
\usepackage{algorithmic}
\usepackage{graphicx}
\usepackage{textcomp}
\usepackage{bm}
\usepackage{graphicx}
\usepackage{float}
\usepackage{ascmac} 
\usepackage{fancybox}
\usepackage{float}
\usepackage{tabularx}
\usepackage{booktabs}
\usepackage{amssymb}
\usepackage{multirow}
\usepackage{cite}
\usepackage{multirow}
\usepackage{siunitx}
\usepackage[normalem]{ulem}
\useunder{\uline}{\ul}{}
\usepackage{url}
\makeatletter
\AtBeginDocument{\DeclareMathVersion{bold}
\SetSymbolFont{operators}{bold}{T1}{times}{b}{n}
\SetMathAlphabet{\mathrm}{bold}{T1}{times}{b}{n}
\SetMathAlphabet{\mathit}{bold}{T1}{times}{b}{it}
\SetMathAlphabet{\mathbf}{bold}{T1}{times}{b}{n}
\SetMathAlphabet{\mathtt}{bold}{OT1}{pcr}{b}{n}
\SetSymbolFont{symbols}{bold}{OMS}{cmsy}{b}{n}
\renewcommand\boldmath{\@nomath\boldmath\mathversion{bold}}}
\makeatother

\def\x{{\bm x}}
\def\w{{\bm w}}
\def\paramssl{{\bm \theta_{\text{ssl}}}}
\def\SSL{{\tt SSL}}
\def\WS{{\tt WS}}
\def\F{{\bm F}}
\def\inRT{\in \mathbb{R}^{T}}
\def\inRdT{\in \mathbb{R}^{D \times T'}}
\def\Fn{\F_n}
\def\Fw{\F^{\tau}}
\def\wn{w^{\tau}_n}
\def\paramw{{\bm \theta}_{\text{w}}^{\tau}}
\def\l{\bm l}
\def\lk{\l^{\tau}}
\def\lhatk{\hat{\l}^{\tau}}
\def\what{\hat{\bm w}}
\def\DOWNSTREAMk{{\tt DS}^{\tau}}

\def\paramdsk{{\bm \theta_{\text{ds}}^{\tau}}}
\def\SE{{\tt SE}}
\def\ASR{{\tt ASR}}
\def\LMFB{{\tt LMFB}}
\def\FE{{\tt FE}}
\def\y{{\bm y}}
\def\shat{\hat{{\bm x}}}
\def\stilde{\tilde{{\bm x}}}
\def\s{{\bm x}}
\def\paramse{{\bm \theta_{\text{se}}}}
\def\paramfe{{\bm \theta_{\text{fe}}}}
\def\paramasr{{\bm \theta_{\text{asr}}}}
\def\L{{\mathcal L}}
\def\Lsnr{\L_{\text{SNR}}}
\def\Lsslmse{\L_{\text{SSL}}}
\def\Lsslmsemt{\L_{\text{SSL-MT}}}

\def\Lasr{\L_{\text{ASR}}}
\def\Lasrm{\L_{\text{ASR-MT}}}
\def\Llmfb{\L_{\text{LMFB}}}
\def\Llmfbm{\L_{\text{LMFB-MT}}}
\def\Lctc{\L_{\text{ctc}}}
\def\Latt{\L_{\text{att}}}
\def\Llmfb{\L_{\text{LMFB}}}

\def\keyFont{\fontsize{8}{11}\helveticabold }
\def\firstAuthorLast{Hiroshi Sato {et~al.}} 
\def\Authors{Hiroshi Sato\,$^{1,*}$, Tsubasa Ochiai\,$^{1}$, Marc Delcroix\,$^{1}$, Takafumi Moriya\,$^{1}$, Takanori Ashihara\,$^{1}$, and Ryo Masumura\,$^{1}$}
\def\Address{$^{1}$NTT Corporation, Yokosuka-shi, Kanagawa, Japan}
\def\corrAuthor{Corresponding Author}

\def\corrEmail{hrs.sato@ntt.com}

\begin{document}
\onecolumn
\firstpage{1}

\title[Generic Speech Enhancement with SSL Loss]{Generic Speech Enhancement with Self-Supervised Representation Space Loss} 

\author[\firstAuthorLast ]{\Authors} 
\address{} 
\correspondance{} 

\extraAuth{}

\maketitle

\begin{abstract}
Single-channel speech enhancement is utilized in various tasks to mitigate the effect of interfering signals. Conventionally, to ensure the speech enhancement performs optimally, the speech enhancement has needed to be tuned for each task. Thus, generalizing speech enhancement models to unknown downstream tasks has been challenging. This study aims to construct a generic speech enhancement front-end that can improve the performance of back-ends to solve multiple downstream tasks. To this end, we propose a novel training criterion that minimizes the distance between the enhanced and the ground truth clean signal in the feature representation domain of self-supervised learning models. Since self-supervised learning feature representations effectively express high-level speech information useful for solving various downstream tasks, the proposal is expected to make speech enhancement models preserve such information. Experimental validation demonstrates that the proposal improves the performance of multiple speech tasks while maintaining the perceptual quality of the enhanced signal.

\tiny
 \keyFont{ \section{Keywords:} Self-supervised learning, Loss function, SUPERB benchmark, Signal denoising, Speech enhancement, Deep learning, Speech recognition} 
\end{abstract}

\section{Introduction}
The recent advancements in machine learning technology have rapidly improved the level of machine comprehension of natural language. In our daily lives, natural language often serves as a medium for communication in the form of speech. To achieve machine understanding on the basis of such spontaneous speech and to enable natural responses to spoken language, studies have extensively explored various speech tasks such as automatic speech recognition (ASR), automatic speaker verification (ASV)~\citep{yang2021superb,tsai2022superb,prabhavalkar2023end,huh2023voxsrc,wani2021comprehensive}. To apply these speech technologies in real-world environments, which often include various types of noise, speech enhancement (SE) is often introduced as the front-end that suppresses such noises as urban bustle, automobile sounds, household clatter from dishes, and the keystrokes and clicks of typing~\citep{delcroix2018single,eskimez2018front,o2021conformer,pandey2021dual,chang2022end,masuyama2022end,lu22c_interspeech,fujita2024noise,kandagatla2020performance}. It has been shown that an SE front-end can improve subsequent speech tasks such as ASR~\citep{kinoshita2020improving}, speech emotion recognition~\citep{avila18_interspeech}, and ASV~\citep{shon19b_interspeech}. In this work, we call the speech enhancement module to enhance input signal for subsequent tasks as ``front-end'', and the subsequent task as ``back-end''.

Besides, the SE front-end is crucial for improving human listening. For instance, SE technology is widely applied in web conferences to remove background noise and facilitate smoother communication. Various research efforts have been devoted to improving the quality of speech recording~\citep{reddy2020interspeech,dubey2022icassp,dubey2024icassp}. Briefly stated, SE front-end is a general technique to improve automatic speech processing technologies and human auditory experiences. 
Hereafter, the back-end tasks related to machine-based speech processing, along with the task aimed at improving human listening, will be collectively referred to as {\it downstream tasks}.

We focus on single-channel SE that aims at extracting a clean speech signal from a monaural recording of noisy speech. Numerous methods have been proposed to date as single-channel SE techniques~\citep{reddy2020interspeech,dubey2022icassp,dubey2024icassp}. SE is broadly divided into two categories: either in the time-frequency (TF) domain~\citep{wang19h_interspeech,vzmolikova2019speakerbeam,hu20g_interspeech,xia2020weighted,hao2021fullsubnet,zhao2022frcrn,ju2023tea} or directly in the time domain~\citep{luo2019conv,delcroix2020improving,pandey2020densely,defossez20_interspeech,wang2021tstnn,sato2024speakerbeam}. Deep complex convolution recurrent network (DCCRN) is one of the major TF domain approaches~\citep{hu20g_interspeech}. DCCRN enhances complex valued TF representation with the backbone network using a recurrent neural network (RNN) and convolutional neural network (CNN) and performs so well that it ranked first in the real-time-track in Interspeech 2020 Deep Noise Suppression (DNS) Challenge~\citep{reddy2020interspeech}. Conv-TasNet is a representative method of the time-domain approach, which directly enhances the input waveform to obtain the enhanced waveform~\citep{luo2019conv}. Initially proposed for speech separation, Conv-TasNet performs superiorly to the ideal TF mask.

However, the improvements in SE performance measures such as signal-to-distortion ratio (SDR) do not necessarily mean that the SE method will improve the performance of downstream tasks when used as the front-end. It is reported that SE sometimes degrades back-end task performance despite the improvement in such performance measures as SDR~\citep{yoshioka2015ntt,chen2018building,fujimoto2019one}. This is because deep neural network (DNN)-based SE tends to generate ``processing artifacts'' due to non-linear transformations, which are detrimental to the subsequent downstream tasks~\citep{iwamoto2022how,sato2022learning,sato2021should}. Since such processing artifacts are unknown to downstream tasks during training, the negative effects of these distortions tend to outweigh the benefits of noise removal. In other words, there is a mismatch between the output of the SE whose training objective (e.g., SDR) attends to remove noise as much as possible while introducing artifacts and the input expected by subsequent tasks, which may be sensitive to unexpected artifacts.

Several methods have been investigated to reduce the mismatch between the SE front-end and the back-end. One approach is to train the back-end model with the enhanced signal, which is output by the SE front-end. This method allows the back-end model to learn, including the distortions present in the enhanced signal, thereby reducing the mismatch~\citep{kinoshita2020improving}. However, this method makes the back-end dependent on a specific front-end enhancement model, which compromises the system's modularity, making it impossible to develop the front-end and back-end independently. Additionally, this approach cannot be applied if the subsequent task involves using API services to solve back-end speech tasks, such as Google's speech-to-text API~\citep{googleapi}, Open AI speech-to-text API~\citep{openaiapi}, or employing off-the-shelf black-box models since we cannot retrain them. Besides, this approach may not be applicable when the back-end model is extremely large, such as large audio-language models (LALMs), where retraining is not feasible.

Another approach is to integrate the SE front-end with the back-end task and jointly optimize them using the training criteria of the latter task~\citep{gao2015joint,wang2016joint,menne2019investigation,subramanian2019speech}. This can mitigate the mismatch between the two models. However, this method has a limitation that the SE front-end becomes specialized for a particular subsequent system, which may prevent it from fully performing when combined with different back-end systems. From a practical point of view, it is ideal to improve the performances of the SE front-end and back-end models separately. Requiring joint training every time one component is updated incurs significant operational costs, which is often not practical.

In this study, we aim to develop a single, generic SE model that is constructed independently of subsequent tasks and can be applied to various downstream applications, including automatic speech understanding technologies and human auditory experiences. The core idea is to train the SE model using a loss function that encompasses speech representations suitable for various downstream tasks. Recently, self-supervised learning (SSL) has been proposed and has proven effective in learning such generic representations. On the basis of this, we propose the SSL Mean Square Error (SSL-MSE) loss, which computes the loss within the SSL representation space. SSL, which is a research field that has gained significant attention in recent years, can learn useful representations without the need for artificially labeled data~\citep{mohamed2022self,liu2021self,jaiswal2020survey}. The proposed SSL-MSE loss is a training criterion to minimize the distance between the ground truth clean signal and enhanced signal in the feature domain that is extracted by the pre-trained SSL model. Since SSL models have been shown to learn effective representations for many downstream tasks, they can capture not only acoustic but also higher-level information such as phonetic or semantic information~\citep{pasad2021layer,dunbar21_interspeech,hsieh21_interspeech}. Thus, we expect that creating a loss term in the SSL domain could guide SE training to preserve or enhance various levels of characteristics of the speech signals that may be required for high-level downstream tasks.

This approach enables not only a single SE model to be applied to multiple tasks but also the SE front-end and the back-end to be independently improved in solving each task, thereby maintaining modularity and allowing each technology to be independently maintained and developed. Additionally, an SE model will be able to be constructed that can improve the performance of downstream tasks that are black-box and cannot be retrained, as well as tasks with a very large number of parameters and for which retraining is impractical.

In this work, we evaluate the effectiveness of the proposal with various back-end tasks. We evaluated speech recognition, speaker verification, and intent classification performance using the SSL model in the pipeline, as well as a speech recognition model using an off-the-shelf Whisper model and a human listening task, i.e., measuring the objective measures of perceptual quality of the enhancement signal by perceptual evaluation of speech quality (PESQ)~\citep{rix2001perceptual} and deep noise suppression mean opinion score (DNSMOS)~\citep{reddy2022dnsmos}.

The main contributions of this paper can be summarized as follows:
\begin{enumerate}
\item We propose a novel training loss that minimizes the distance between enhanced and clean speech signals in the SSL representation space. The proposed SSL-MSE loss transfers the generalizability of SSL models over various downstream tasks into SE models.
\item We perform an extensive evaluation on multiple downstream tasks, showing that models trained with the proposed SSL-MSE loss outperform vanilla SE models as well as models trained on different representation spaces such as log mel-filter bank (LMFB) or ASR loss.
\end{enumerate}

While our approach shares the idea of utilizing SSL representations with the prior work proposed by Hsieh et al.~\citep{hsieh21_interspeech}, there are key differences in both the formulation and objectives. Hsieh et al. primarily aim to improve perceptual quality for human listeners and proposed phonefortified perceptual loss (PFPL) defined on the final-layer output of wav2vec 2.0. On the other hand, this study focused on building a generic SE model that is effective across a wide range of downstream tasks. To obtain such versatility, we proposed the SSL-MSE loss that incorporates weighted representations from the latter half of the SSL layers. Additionally, to show the versatility for various downstream tasks, we have conducted a comprehensive experiments using a variety of SSL models and downstream task evaluations. These distinctions in design and effect underline the novelty and utility of our approach.


This paper extends our previous short conference paper, where we introduced the basic concept of SSL-MSE and showed that SSL-MSE loss could help downstream models that utilize SSL models for feature extraction~\citep{sato23_interspeech}. 
In this work, we provide a more detailed explanation of SSL-MSE, adding more context. In addition, we include new experimental results showing that it can also benefit models that do not rely on SSL features, thereby achieving a truly generic model. To evaluate the generalizability over downstream tasks, we extend experimental validations to include objective measures of perceptual quality, such as PESQ~\citep{rix2001perceptual} and DNSMOS~\citep{reddy2022dnsmos}, which are known to correlate with human subjective evaluation, as well as ASR performance evaluation using the Whisper model~\citep{radford2023robust}. Furthermore, we examine a combination of SSL-MSE training with an observation adding (OA) post-processing technique to further reduce the effect of the distortion generated by the SE.

The rest of the paper is organized as follows. In Section 2, we will discuss the related works. In Section 3, we explain the conventional method, and in Section 4, we explain the proposed approach. In Section 5, we detail the experimental validation. Section 6 concludes the paper.

\section{Related Works}
\subsection{SSL models}
\label{sec:ssl}
SSL in speech is a technology for learning powerful representations. For speech SSL models, various techniques have been proposed~\citep{tera,mockingjay,schneider19_interspeech,BaevskiSA20,chang2022distilhubert,baevski2020wav2vec,hsu2021hubert,chen2022wavlm,chen2023joint, yadav24b_interspeech, DBLP:conf/iclr/ShiIMKS24}, among which wav2vec 2.0~\citep{baevski2020wav2vec}, HuBERT~\citep{hsu2021hubert}, and WavLM~\citep{chen2022wavlm} are some of the most widely applied approaches.

The SSL model is trained to extract a time series of feature representation from a single-channel audio, which is useful for many types of downstream tasks that can be learned on the representation with a limited amount of paired data. 
The task of training the upstream SSL model is called the pre-text task, which differs for each type of SSL model. 

For example, wav2vec 2.0~\citep{baevski2020wav2vec} masks parts of the input speech in the latent space and solves a contrastive prediction task over quantized representations, encouraging the model to learn the structure and patterns of speech without labeled data.
Another key aspect of wav2vec 2.0 is its use of quantization. After initial processing, the audio is transformed into quantized representations by mapping the continuous audio features into a finite set of codebook entries. Quantization enables the model to represent audio with a smaller, more manageable set of discrete units, which captures essential information about the speech content while reducing redundancy. 
This discrete representation is used as the target for the model’s predictions, allowing it to focus on the fundamental components of speech rather than unnecessary detail.
HuBERT~\citep{hsu2021hubert} and WavLM~\citep{chen2022wavlm} are trained using the BERT-like masked prediction task~\citep{devlin2018bert} on the target label generated in an offline clustering step.
By masking of the input, HuBERT is trained to predict these hidden units for the masked frames, helping it capture essential patterns in speech structure and phonetic information.
Of particular note, WavLM is made robust to noise and interference speakers through introducing a denoising task into the pre-text task. Specifically, DNS noise and interfering speech are added to the input audio, and WavLM is trained to ignore these signals. Although a prior study showed that WavLM was relatively robust to noise ~\citep{masuyama2022end}, Chung {\it et al.} revealed that the noise robustness of SSL models could be further improved by introducing a single-channel SE front-end~\citep{chang2022end}. 
Another approaches have been proposed for improving the noise robustness of SSL system by performing speech enhancement directly in the SSL feature domain~\citep{ali2023direct, ali2022enhancing}.

As a result of the training on the pre-text task, SSL upstream models function as a universal feature extractor that can generally be applied for various downstream tasks~\citep{yang2021superb,tsai2022superb}. Note that the feature extracted from the SSL model is calculated as the weighted sum over the outputs from the transformer layers of SSL models where different weights are used for different downstream tasks. This reveals that SSL models can capture rich speech representation through their layers. The proposed SSL-MSE aims to transfer the generalizability of the SSL model into an SE model by using the loss term calculated on the SSL representations.

\subsection{Processing artifacts}
Another deeply related research topic is the processing artifacts generated by SE. DNN-based SE generates processing artifacts that are detrimental to subsequent tasks. As mentioned in Introduction, joint training is one way to mitigate artifacts, but it implies creating SE specific for each downstream task, which is not practical. A promising approach to mitigate the mismatch caused by the processing artifact is OA post-processing, which interpolates the enhanced and observed signals as the input signal to the back-end~\citep{ochiai2024rethinking}. OA improves the signal-to-artifact ratio instead of increasing the signal-to-noise ratio (SNR). The appropriate ratio of OA improves back-end ASR performance since artifacts tend to be more detrimental than noise to ASR. In this work, we investigate the combination of the proposed SSL-MSE and the OA approach. 

\section{Fundamental System Overview}

\subsection{Self-supervised learning}
There are two major ways to apply SSL models to downstream tasks, either as a fixed feature extractor, or permitting retraining of their parameters~\citep{mohamed2022self}. Following previous studies~\citep{yang2021superb,chang2022end,masuyama2022end}, we adopted the former approach, i.e., freezing the parameters of the SSL model $\paramssl$ during the training of the downstream models and using the SSL upstream model as a feature extractor, because SSL models are usually very large and thus fine-tuning the SSL models for every downstream task is computationally too intensive. The downstream task-specific additional layers are relatively small DNNs, which we call the downstream model.

Formally, we can write the feature extraction process with the SSL model as follows~\citep{yang2021superb}:
\begin{align}
 \label{eq:ssl}
 \F_{1:N} = \SSL(\x;\paramssl),
\end{align}
where $\x \inRT$ denotes the monaural input signal in raw waveform with $T$ samples, $\F_n \inRdT$ denotes the time series of extracted features obtained from the $n$-th layer of the SSL model, and $\F_{i:j}$ denotes $\F_{i:j} = [ \F_i, \F_{i+1}, ..., \F_j]$. $N$ denotes the number of layers of the SSL model, and $\paramssl$ denotes the learnable parameters of the SSL model. $T'$ and $D$ denote the number of frames and dimensions of the extracted features, respectively. The learnable parameters, $\paramssl$ are trained with the pre-text task.

An effective way to apply the SSL features across various types of downstream tasks is to use the weighted sum of the embeddings from different layers in the SSL model as the input feature of the downstream model~\citep{chen2022wavlm, yang2021superb}. The process of the downstream model for a task $\tau$ can be written as follows:
\begin{align}
 \label{eq:weighting}
 \Fw & = \WS(\F_{1:N};\paramw)
 \equiv \sum_{n=1}^{N} \wn\Fn, \\
 \label{eq:downstream}
 \lhatk & = \DOWNSTREAMk(\Fw;\paramdsk),
\end{align}
where $\WS(\cdot)$ is the weighted sum function, $\paramw = [w_1^{\tau},...,w_N^{\tau}]$ are the weights for latent representations obtained from each layer, which are learnable parameters, and $\lhatk$ is the estimation result attained by the downstream model. $\DOWNSTREAMk(\cdot)$ denotes the downstream model for the task $\tau$ whose learnable parameters are $\paramdsk$. Since SSL models capture varying levels of representation across layers, the latent representation from each layer $\Fn$ retains information at different granularities. By using the weighted sum of these representations as input to the downstream model, the downstream model is expected to be able to effectively solve a variety of tasks. The learnable parameters $(\paramw,\paramdsk)$ are jointly optimized by using task-specific paired data $(\lk, \lhatk)$ where $\lk$ is the ground truth label, while the upstream model $\paramssl$ is frozen.

\subsection{Speech Enhancement}
We investigate single-channel neural-based SE as the front-end. Let $\y \inRT$ and $\shat \inRT$ be the noisy observations and enhanced signals, respectively. The enhancement process can be denoted as follows:
\begin{align}
 \label{eq:se}
 \shat = \SE(\y; \paramse).
\end{align}

$\paramse$ denotes the learnable parameters of the SE model. To train the SE model, paired data of noisy observation $y$ and ground truth clean source $\s \inRT$ are commonly prepared. The SE model is commonly optimized by minimizing the distance between enhanced signal $\shat$ and the ground truth clean source $\s$. Specifically, we adopt the scale-dependent SNR loss $\Lsnr(\cdot)$ as the distance measure, which is defined as follows:

\begin{align}
 \label{eq:sdsnr}
 \Lsnr = -10\log_{10}\frac{\lVert\shat\rVert^2}{\lVert\s-\shat\rVert^2}.
\end{align}
Since the clean source is not available for real-recorded noisy speech, simulated noisy speech $\y$ is generated on the basis of the speech $\s$ and noise recordings for SE training~\citep{vincent2006performance}.

\subsection{Combination of SE front-end with back-end tasks}
\begin{figure}[tpb]
 \begin{center}
 \includegraphics[width=\hsize]{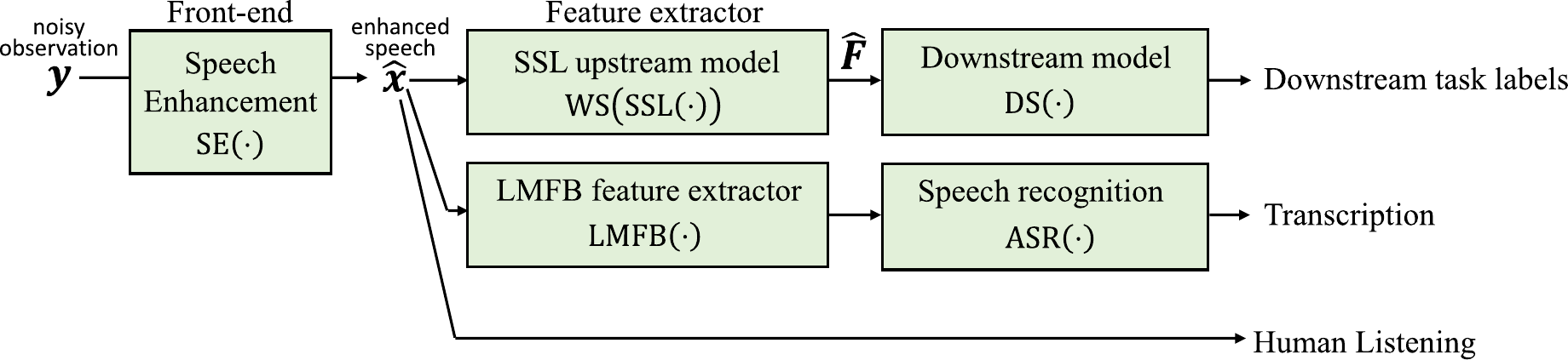}
 \end{center}
\caption{Overview of the combination of SE front-end with back-end tasks. The backend task can either be 1) a speech processing task with SSL upstream and downstream models, 2) a speech processing task without SSL models, or 3) a human listening task.}
 \label{fig:overview}
\end{figure}

The introduction of an SE front-end can improve the performance of subsequent tasks. The subsequent task can either be 1) a speech processing task with SSL upstream and downstream models, 2) a speech processing task without SSL models, or 3) a human listening task. The overview of the combination of SE front-end with back-end tasks is shown in Figure~\ref{fig:overview}. Let the feature extraction process be $\FE(\cdot)$, the general pipeline to solve a downstream task with SE can be expressed as follows:
\begin{align}
 \label{eq:fe}
 \lhatk = \DOWNSTREAMk(\FE(\SE(\y; \paramse); \paramfe); \paramdsk).
\end{align}
where $\paramfe$ denotes the learnable parameters of the feature extractor.

\subsubsection{Back-end with SSL models}
For a speech understanding task with SSL models, the feature extractor in Eq.~\eqref{eq:fe} is an SSL upstream model and weighted sum function. The whole pipeline can be expressed as follows~\citep{yang2021superb}:
\begin{align}
 \label{eq:sslpipeline}
 \lhatk = \DOWNSTREAMk(\WS(\SSL(\SE(\y; \paramse);\paramssl);\paramw);\paramdsk).
\end{align}

Previous research reports that to benefit from SE, it is necessary to fine-tune SE that is trained independently from the SSL pipeline~\citep{masuyama2022end} to mitigate the mismatch between the front-end and the back-end. Specifically, to mitigate this mismatch, $(\paramse,\paramw,\paramdsk)$ are jointly optimized by using downstream task paired data while freezing SSL model $\paramssl$~\citep{masuyama2022end}. However, the fine-tuning of the downstream task makes the SE front-end task-specific, making it infeasible to share a single general SE model for multiple downstream tasks, thus incurring development costs. Besides, when considering more than a few downstream tasks, optimizing the SE front-end and back-end model jointly for every downstream task may be computationally demanding.

\subsubsection{Back-end without SSL models}
For the back-end tasks without SSL models, the feature extractor in Eq.~\eqref{eq:fe} is typically an LMFB feature extractor. The whole processing pipeline can be expressed as follows:
\begin{align}
 \label{eq:asrse}
 \lhatk = \DOWNSTREAMk(\LMFB(\SE(\y; \paramse));\paramdsk).
\end{align}
For systems that do not use SSL, the mismatch between the SE front-end and back-end is also a challenge. As a way to mitigate the mismatch, LMFB feature domain loss has been proposed that calculates the loss term on the LMFB feature domain~\citep{wangvoicefilter}. Preceding works show the effectiveness of LMFB feature domain loss in improving the back-end ASR task with SE. In this work, we examine the multitask loss of LMFB feature domain loss and SNR loss. Formally, the multitask loss $\Llmfbm$ is expressed as follows~\citep{wangvoicefilter}:
\begin{align}
 \label{eq:lmfbloss}
 \Llmfbm =& \Llmfb(\shat, \x)+ \alpha \Lsnr(\shat, \x), \\
 \Llmfb =& \lVert \LMFB(\shat) - \LMFB(\x) \rVert_{\hspace{-1pt}F}^2/M
\end{align}
where $\lVert\cdot\rVert_{\hspace{-1pt}F}$ represents Frobenius norm and $\LMFB(\cdot)$ represents the LMFB feature extraction. $\alpha$ represents a multitask weight. $M$ represents the number of elements of LMFB feature $\LMFB(\x)$.

Another way to mitigate the mismatch is to train SE on the loss function that minimizes the distance between enhanced and clean signals on the output of the back-end task. Considering ASR as an example and letting $\ASR(\cdot)$ be the process of the ASR, the ASR result is obtained from a noisy observation via SE as follows:
\begin{align}
 \label{eq:asrse}
 \what = \ASR(\LMFB(\SE(\y; \paramse));\paramasr),
\end{align}
where $\paramasr$ and $\what$ indicate learnable parameters of the ASR model and ASR result, respectively. The ASR output basis loss minimizes the distance between the ASR result of the enhanced signal $\what$ and ground truth transcription $\w$ to optimize the SE model $\paramse$ while freezing the ASR model $\paramasr$. We hereafter refer to this training criterion as ASR loss. We examine the SE training with the multitask loss of the ASR loss and SNR loss as follows:
\begin{align}
 \label{eq:asrloss}
 \Lasrm =& \Lasr(\what, \w)+ \alpha \Lsnr(\shat, \x), \\
 \Lasr =& \lambda \Lctc(\what, \w) + (1-\lambda) \Latt(\what, \w), 
\end{align}
where $\alpha$ is the multitask weight controlling the impact of SNR loss. $\Lctc$ and $\Latt$ optimize connectionist temporal classification (CTC)~\citep{Graves2006} decoder and attention-based decoder, respectively. Note that $\Latt$ is a cross-entropy loss. $\lambda$ is a hyperparameter controlling the balance of CTC and attention loss.

\subsubsection{Human perception as back-end task}
We consider human perception as a back-end task. In this case, there is no feature extraction $\FE(\cdot)$ or downstream model $\DOWNSTREAMk(\cdot)$, as the goal is to produce an enhanced signal, and thus the whole system can be expressed as Eq.\eqref{eq:se}.

\section{Proposed Method}
\label{sec:proposed}

\begin{figure}[tpb]
 \begin{center}
 \includegraphics[width=\hsize]{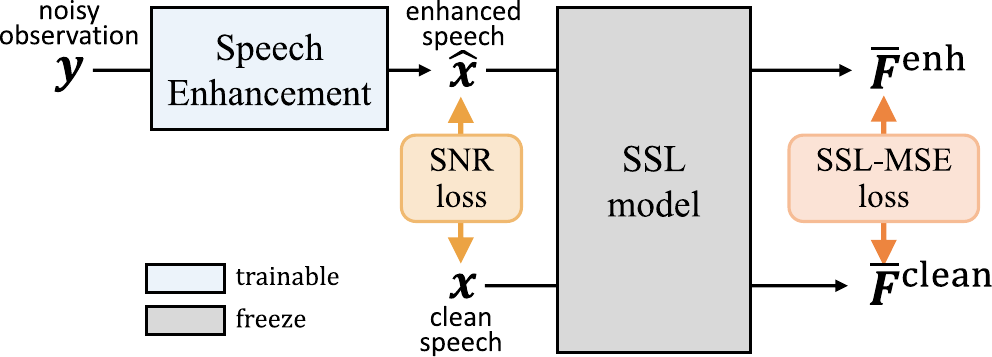}
 \end{center}
\caption{Overview of the proposed SSL-MSE loss. Proposed SSL-MSE loss calculate the distance between enhanced and clean signals on the feature domain extracted by the SSL model.}
 \label{fig:framework}
\end{figure}

To realize an SE that can be applicable to multiple tasks, we propose SSL-MSE. Figure~\ref{fig:framework} shows the general framework of the proposed SSL-MSE training. SSL-MSE loss calculates the distance between enhanced and clean signals on the SSL model's feature domain. Since SSL models have been shown to learn effective representations for many downstream tasks, SSL-MSE can make SE more optimal for the front-end of multiple tasks. More specifically, SSL models are reported to capture not only acoustic but also higher-level information such as phonetic or semantic information~\citep{pasad2021layer,dunbar21_interspeech,hsieh21_interspeech}, and thus a loss term in the SSL domain could guide SE training to preserve or enhance various levels of characteristics of the speech signals that may be required for high-level downstream tasks. Formally, SSL-MSE $\Lsslmse$ between enhanced speech $\shat=\SE(\y; \paramse)$ and clean source $\s$ is calculated as the MSE between SSL features extracted from these signals as follows:
\begin{align}
 \label{eq:SSL-MSE}
 \Lsslmse &= \lVert\overline{{\bm F}}^{\text{enh}} - \overline{{\bm F}}^{\text{clean}}\rVert_{\hspace{-1pt}F}^2/(DT') \\
 \overline{{\bm F}}^{\text{enh}} &= \sum_{n=1}^{N}\Tilde{w}_n\F_n^{\text{enh}}, \\
 \overline{{\bm F}}^{\text{clean}} &= \sum_{n=1}^{N}\Tilde{w}_n\F_n^{\text{clean}}, \\
 \F_{1:N}^{\text{enh}} &= \SSL(\SE(\y; \paramse);\paramssl), \\
 \F_{1:N}^{\text{clean}} &= \SSL(\s;\paramssl),
\end{align}
where $\tilde{\bm{w}} = [\tilde{w}_1,...,\tilde{w}_N]$ represents the layer weight for the latent feature obtained from each layer of the SSL model. The layer weight is a hyperparameter that controls which layer output SSL-MSE focuses on. Since SSL models capture different levels of representations in their different layers, the latent representation from each layer retains information at different levels of granularity. Thus, applying SSL loss over multiple layers may be advantageous as it ensures that the SE module captures necessary information. Specifically, we adopt a weight that uniformly emphasizes the features outputted from the latter half of the layers to preserve higher-level representations, as the lower-level ones are expected to be captured in combination with SNR loss. Formally, the weight is expressed as follows:
\begin{align}
 \tilde{w}_1, ... ,\tilde{w}_{\lfloor 2/N \rfloor} =& 0, \\
 \tilde{w}_{\lfloor 2/N \rfloor+1}, ..., \tilde{w}_N =& \frac{1}{N-\lfloor 2/N \rfloor},
\end{align}
where $\lfloor \cdot \rfloor$ represents the floor function. Although it is possible to calculate the loss on each layer and sum the losses instead of calculating the loss on the weighted sum of the representations, we adopted the latter way to maintain consistency with the feature extraction process of the SSL pipeline as in Eq.~\eqref{eq:sslpipeline}.

The model parameter $\paramse$ is optimized by minimizing the multitask loss function $\Lsslmsemt$ expressed as follows:
\begin{align}
 \label{eq:overallloss}
 \Lsslmsemt = \Lsslmse + \alpha\Lsnr,
\end{align}
where $\alpha$ denotes the multitask weight for the SNR loss. Note that the parameters of the SSL model $\paramssl$ are frozen during the SSL-MSE training.

By developing an SE model that is trained independently from the back-end tasks and is applicable to multiple tasks, we can independently improve both the SE front-end and the back-end systems for each specific task. This approach preserves modularity, allowing each component to be independently maintained and developed. Furthermore, this methodology not only facilitates the use of a single SE model across various tasks but also enables SE models to be constructed that can improve the performance of downstream tasks, even when those tasks are black-box systems that cannot be retrained. It is also beneficial for tasks with a very large number of parameters, where retraining would be impractical.

To further reduce the mismatch between SE front-end and the back-end, we investigate the combination of SSL-MSE training with OA post-processing~\citep{ochiai2024rethinking}. OA post-processing interpolates enhanced and input signals in the waveform signal domain as the final output of the SE. The distortion generated by the SE can be decreased by adding the observation signal, which can also mitigate the mismatch between front-end and back-end. With OA, the input of the back-end task $\stilde$ is expressed as follows:
\begin{align}
 \label{eq:oa}
 \stilde = \beta \y + (1 - \beta) \shat,
\end{align}
where $\beta$ is the OA ratio. Note that OA is implemented as a post-processing step for the enhanced signal, and thus depending on the downstream task, it is possible to modify the adding ratio of OA or omit its processing.

\section{Experiments}
\subsection{Experimental setup}
\subsubsection{Speech enhancement front-end}
The SE model was trained on simulated mixtures of speech and noise data. We used LibriSpeech~\citep{panayotov2015librispeech} for speech recordings and DNS noise~\citep{reddy2021interspeech} for noise recordings sampled at 16 kHz. The number of noisy observations were 100,000 and 5,000 for the training and development sets, respectively. The noise was added at SNR values randomly sampled from -3 to 20 dB. We chose DNS noise to create a fair comparison of the SSL pipeline with/without the SE module, as DNS noise is also used in training the WavLM models. We adopted Conv-TasNet as the SE front-end module, which converts noisy raw-waveform audio into enhanced raw-waveforms in a time-domain, end-to-end processing manner~\citep{luo2019conv}. In accordance with the setup adopted in~\citep{e3net}, we set the hyperparameters to ${\rm N}\hspace{-1mm}=\hspace{-1mm}4096$, ${\rm L}\hspace{-1mm}=\hspace{-1mm}320$, ${\rm B}\hspace{-1mm}=\hspace{-1mm}256$, ${\rm R}\hspace{-1mm}=\hspace{-1mm}4$, ${\rm X}\hspace{-1mm}=\hspace{-1mm}8$, ${\rm H}\hspace{-1mm}=\hspace{-1mm}512$ and ${\rm P}\hspace{-1mm}=\hspace{-1mm}3$ following the notation in \citep{luo2019conv}. The SSL-MSE loss was introduced after pre-training the model with conventional SNR loss to speed up the conversion. The initial learning rate for the pre-training with SNR loss was set to 5e-4 and was multiplied by 3/4 if the loss on the development set did not decrease for 2 epochs. For optimization, we adopted the Adam optimizer \citep{kingma2014adam}. The models were pre-trained for 100 epochs. If we apply OA, we adopt adding ratio $\beta$ of 0.1 and 0.5.

\subsubsection{Multitask loss}\label{sec:multitaskloss}
We investigate three types of multitask losses: SSL-MSE loss, ASR loss, and LMFB loss. Each multitask loss training is applied as the fine-tuning process: we first pretrained the SE model with SNR loss and then fine-tuned it with SSL-MSE loss with the initial learning rate of 1e-4 and for up to 50 epochs. We tested the SNR weight $\alpha$ within $\{0,0.0001,0.001,0.01,0.1,1,10\}$. The comparison is shown in Figure \ref{fig:comparison}. 

\noindent \textbf{SSL-MSE loss:}
To calculate SSL-MSE loss in Eq.~\eqref{eq:overallloss}, we tested four off-the-shelf SSL models: WavLM {\sc Base+}, WavLM {\sc Large}, wav2vec 2.0 {\sc Base}, and Hubert {\sc Base} models. SSL models were frozen during SSL-MSE training. WavLM {\sc Base+} and WavLM {\sc Large} share the same training data; however, WavLM Large has more than three times the number of parameters compared to WavLM Base+. 
Wav2Vec 2.0 and HuBERT are not trained to be robust to noise; their pre-training procedures do not involve the explicit addition of noise. In contrast, WavLM {\sc Base+} and WavLM {\sc Large}~\citep{chen2022wavlm} are pre-trained with noise augmentation to enhance robustness.

\noindent \textbf{ASR loss:}
ASR loss in Eq.~\eqref{eq:asrloss} is calculated by a pre-trained transformer-based hybrid CTC/Attention ASR model~\citep{watanabe2017hybrid,karita2020ctctransformer}, which consists of 12 conformer encoder blocks and 6 transformer decoder blocks. The model was trained following the LibriSpeech training recipe from ESPnet~\citep{watanabe2018espnet}, an open-source toolkit for end-to-end speech processing. $\lambda$ in Eq.~\eqref{eq:asrloss} was set as 0.3. The ASR model is frozen during the ASR loss training.
 
\noindent \textbf{LMFB loss:}
For calculating LMFB loss in Eq.~\eqref{eq:lmfbloss} we adopt LMFB feature extractor of 80 mel bins, 400 sample window length, and 200 sample window shift. 

\subsubsection{Evaluation details}
To evaluate the generalizability of the SE model trained with SSL-MSE loss, we evaluated the performance of SE in terms of the following aspects: 1) SDR~\citep{le2019sdr}, 2) objective measures of subjective quality, 3) the performance of SSL downstream tasks, and 4) Whisper ASR performance. objective measures of subjective quality is evaluated on PESQ and DNSMOS P.835~\citep{reddy2022dnsmos}. SDR, PESQ, and DNSMOS are evaluated on the 3,000 simulated mixtures of LibriSpeech speech recordings and DNS noise where SNR values are randomly selected from 0 to 10 dB.

\noindent \textbf{SDR:}
SDR is evaluated on the 3,000 simulated mixtures of LibriSpeech speech recordings and DNS noise where SNR values are randomly selected from 0 to 10 dB.

\noindent \textbf{Objective measures of perceptual quality:}
Perceptual quality is evaluated using objective measures, PESQ and DNSMOS P.835~\citep{reddy2022dnsmos}, which are known to correlate with human subjective evaluations using the same dataset as SDR evaluation.

\noindent \textbf{SSL downstream tasks:}
To evaluate the performance in SSL downstream tasks, we prepared the SSL pipeline to solve three downstream tasks (automatic speech recognition (ASR), automatic speaker verification (ASV), and intent classification (IC)) following the procedure explained in Section~\ref{sec:ssl}. These performance are evaluated in terms of word error rate (WER), equal error rate (EER), and accuracy (Acc), respective. We implemented the SSL pipeline using the S3PRL toolkit~\citep{yang2021superb,s3prl}. As the SSL model for the inference, we tested WavLM {\sc Base+}, WavLM {\sc Large}, wav2vec 2.0 {\sc Base}, and Hubert {\sc Base}. The performance of SSL pipeline was evaluated on the noisy version of the SUPERB test sets for each task by adding DNS noise to the original recordings at SNR values randomly sampled from 0 to 10 dB. Since DNS noise contains a wide variety of the noise types and we use different samples for training and testing, we can verify robustness to unseen noise although we adopted DNS noise in both training and evaluation.

We tested two setups for downstream model training: 1) {\it official} setup where downstream models were trained with SUPERB official training data~\citep{yang2021superb} that is relatively `clean' speech data, in accordance with the S3PRL SUPERB recipe~\citep{s3prl}, and 2) {\it noise-robust} setup where DNS noise was added to SUPERB official downstream training data at SNR values randomly sampled from -3 to 20 dB. The results gained in the noise-robust setup are shown in Table~\ref{tab:robust}; for the other experiments, we adopt the official setup. The SE front-end was not applied while training the downstream model in noise-robust setups and in official setups. The downstream models were prepared for each downstream task for each SSL model.

\noindent \textbf{Whisper ASR:}
Whisper ASR performance is measured with an off-the-shelf Whisper {\sc medium} model on the simulated mixture of LibriSpeech speech recordings and DNS noise at the SNR values randomly sampled from 0 to 10 dB. The Whisper ASR model is only used for evaluation and not for ASR-loss training. Note that the Whisper model uses LMFB as a feature extraction and does not use SSL models.

\subsection{Results and Discussion}
\subsubsection{The effect of SSL-MSE}
\begin{table}[tpb]
\centering
\sisetup{detect-weight,mode=text}
\renewrobustcmd{\bfseries}{\fontseries{b}\selectfont}
\renewrobustcmd{\boldmath}{}
\newrobustcmd{\BOLD}{\bfseries}
\caption{Performance evaluation of the conventional and proposed systems. For the evaluation of SSL downstream tasks, we used the SSL pipeline with WavLM {\sc Base+} model. The column SSL-MSE loss indicates the type of SSL model used for calculating SSL-MSE loss. WavLM B+ stands for WavLM {\sc Base+}, and WavLM L stands for WavLM {\sc Large}. The SNR multitask loss weight $\alpha$ for SSL-MSE loss training is fixed to 0.1. Values in bold indicate the results of t-tests conducted at a 5\% significance level, showing no significant difference from the highest-performing results. N/A indicates not applicable for each term.}
\label{tab:main}
\resizebox{\textwidth}{!}{ 
\begin{tabular}{@{}lccc|S[table-format=2.1]|S[table-format=1.2]S[table-format=1.2]S[table-format=1.2]S[table-format=1.2]|S[table-format=2.1]S[table-format=2.1]S[table-format=2.1]|S[table-format=2.1]@{}}\toprule
& \multirow{3}{*}{\begin{tabular}[c]{@{}c@{}}SE \\ method \\ or its \\ training \\ method \end{tabular}} & \multirow{3}{*}{\begin{tabular}[c]{@{}c@{}}SSL model\\ for SSL-MSE \\ loss\end{tabular}} & \multirow{3}{*}{\begin{tabular}[c]{@{}c@{}}OA \\ ratio\end{tabular}} & {\multirow{3}{*}{\begin{tabular}[c]{@{}c@{}}SDR↑\\ \text{[dB]}\end{tabular}}} & \multicolumn{4}{c|}{Human perception} & \multicolumn{3}{c|}{SSL downstream tasks} & {Whisper} \\ \cmidrule(lr){6-9} \cmidrule(lr){10-12}
 & & & & & {\multirow{2}{*}{PESQ↑}} & \multicolumn{3}{c|}{DNSMOS} & {ASR} & {ASV} & {IC} & {ASR} \\ \cmidrule(lr){7-9}
 & & & & & & {SIG↑} & {BAK↑} & {OVRL↑} & {\begin{tabular}[c]{@{}c@{}}WER↓\\ \text{[\%]}\end{tabular}} & {\begin{tabular}[c]{@{}c@{}}EER↓\\ \text{[\%]}\end{tabular}} & {\begin{tabular}[c]{@{}c@{}}Acc↑\\ \text{[\%]}\end{tabular}} & {\begin{tabular}[c]{@{}c@{}}{WER↓}\\ \text{[\%]}\end{tabular}} \\ \midrule
(a1) & clean & & & {-} & {-} & 4.02 & 4.02 & 3.57 & 5.6 & 4.4 & 98.8 & 3.9 \\
(a2) & no process & & & 3.7 & 1.27 & 2.88 & 1.86 & 1.99 & 17.1 & 10.9 & 67.4 & 9.3 \\ \midrule
(b1) & SNR loss \citep{luo2019conv} & N/A & N/A & \BOLD 15.7 & 2.27 & 3.73 & \BOLD 4.21 & \BOLD 3.47 & 14.6 & 8.6 & 84.8 & 12.2 \\ 
(b2) & & N/A & 0.1 & 14.5 & 2.01 & 3.82 & 2.88 & 2.93 & 12.2 & 7.8 & 87.8 & 10.3 \\ \midrule
 (c1) & SSL-MSE loss & WavLM B+ & N/A & \BOLD 15.8 & \BOLD 2.35 & 3.74 & 4.18 & \BOLD 3.46 & 12.9 & 8.1 & 87.7 & 11.1 \\
 (c2) & & WavLM L & N/A & 15.6 & \BOLD 2.34 & 3.82 & 3.94 & 3.42 & 11.3 & 7.4 & 88.6 & 9.6 \\ 
(c3) & & WavLM L & 0.1 & 14.3 & 1.97 & \BOLD 3.85 & 2.83 & 2.92 & \BOLD 10.9 & \BOLD 7.2 & \BOLD 88.7 & 
\BOLD 9.1 \\ \bottomrule
\end{tabular}
}
\end{table}

Table \ref{tab:main} compares the performances of SE models trained with conventional SNR loss training and proposed SSL-MSE loss. The SNR multitask loss weight $\alpha$ is fixed to 0.1. 
Comparing (a1) and (a2), the addition of noise to the test sets has a substantial impact on downstream performance. 
For instance, the Whisper ASR system exhibits a WER degradation from 3.9\% in (a1) to 9.3\% in (a2) due to noise addition. This finding aligns with the observations in the original Whisper paper~\citep{radford2023robust}, particularly in Figure 5, which demonstrates that although Whisper models are trained to be robust to noise, they are not entirely impervious to its influence.
The introduction of the baseline SE front-end (b) improves each performance metric, except Whisper ASR WER. Since the Whisper ASR model itself is trained to be robust to the noise, the negative effect of the mismatch between the SE front-end and back-end seems to outweigh the positive effect of the reduction of the noise.

The systems (c1) and (c2) shows the effect of introducing the SSL-MSE loss, with a different SSL model used for calculating SSL-MSE loss, i.e., WavLM {\sc Base plus} (c1) and WavLM {\sc Large} (c2). As for SDR and objective measures of subjective quality, the proposed systems (c1) and (c2) perform equivalently to the baseline system (b). More specifically, the proposed systems improve PESQ and DNSMOS SIG values and degrade DNSMOS BAK and DNSMOS OVRL values. The increase in the BAK value at the expense of the SIG value suggests that while the noise suppression effect is somewhat diminished, the harmful distortion that affects subsequent processing is reduced. The performances of SSL downstream tasks and Whisper ASR were substantially improved by introducing SSL-MSE loss. The improvement is larger when SSL-MSE loss is calculated with the WavLM {\sc Large} model shown in (c2). The effect of the type of SSL model on the performance will be discussed in detail in Section \ref{sec:cross}. Compared with the baseline system (b), the proposed system (c2) improved the WER of the SSL downstream ASR task by 22 \%, the EER of the ASV task by 14 \%, and the Acc of the IC task by 3.8 \%. Based on the observation above, the introduction of SSL-MSE loss improves the performance as the front-end of the subsequent speech understand tasks while maintaining the human perceptual quality, and thus it increases the generalizability of the SE for multiple tasks. Although the proposed system (c2) improved the WER of the Whisper ASR model by 21 \% compared with the baseline system (b), it still performs worse than the system without SE (a2).

System (c3) demonstrates the combination of SSL-MSE loss and OA with the addition ratio $\beta = 0.1$. The addition of the observed signal demonstrates the best SSL downstream task performance and the Whisper ASR performance among all systems. Improving the performance of a noise-robust ASR model with an SE model trained separately from ASR model is generally challenging \citep{kinoshita2020improving}. Despite that, the proposed SE front-end in the system (c3) reduces the WER for the Whisper ASR model, which is trained to be robust to noise. In summary, the combination of OA with SSL-MSE loss training can further improve specific downstream task performance. Note that since the OA can be applied as a post-processing, the decision of whether or not to apply OA can change depending on the type of the downstream tasks. Thus, we can omit OA for human listening tasks using the same SE model.

While the OA technique alone, as used in system (b2), already shows promising generalization across downstream tasks without requiring complex training procedures, a comparison with system (c2), which uses SSL-MSE loss without OA, reveals that the proposed SSL-MSE training provides consistent performance gains. Specifically, system (c2) achieves better results in human perceptual quality (DNSMOS OVRL: 3.42 vs. 2.93), SSL-based ASR (WER: 11.3\% vs. 12.2\%), and Whisper ASR (WER: 9.6\% vs. 10.3\%) compared to system (b2). These results indicate that, although OA is effective as a lightweight post-processing method, the SSL-MSE loss enables the SE model itself to produce higher-quality outputs that are more suitable for both human perception and machine-based downstream tasks. This suggests that the benefit of SSL-based training is not merely complementary to OA but essential for achieving a more universally robust enhancement system.

\subsubsection{Comparison with other training criteria}
\begin{figure}[tpb]
 \begin{center}
 \includegraphics[width=\hsize]{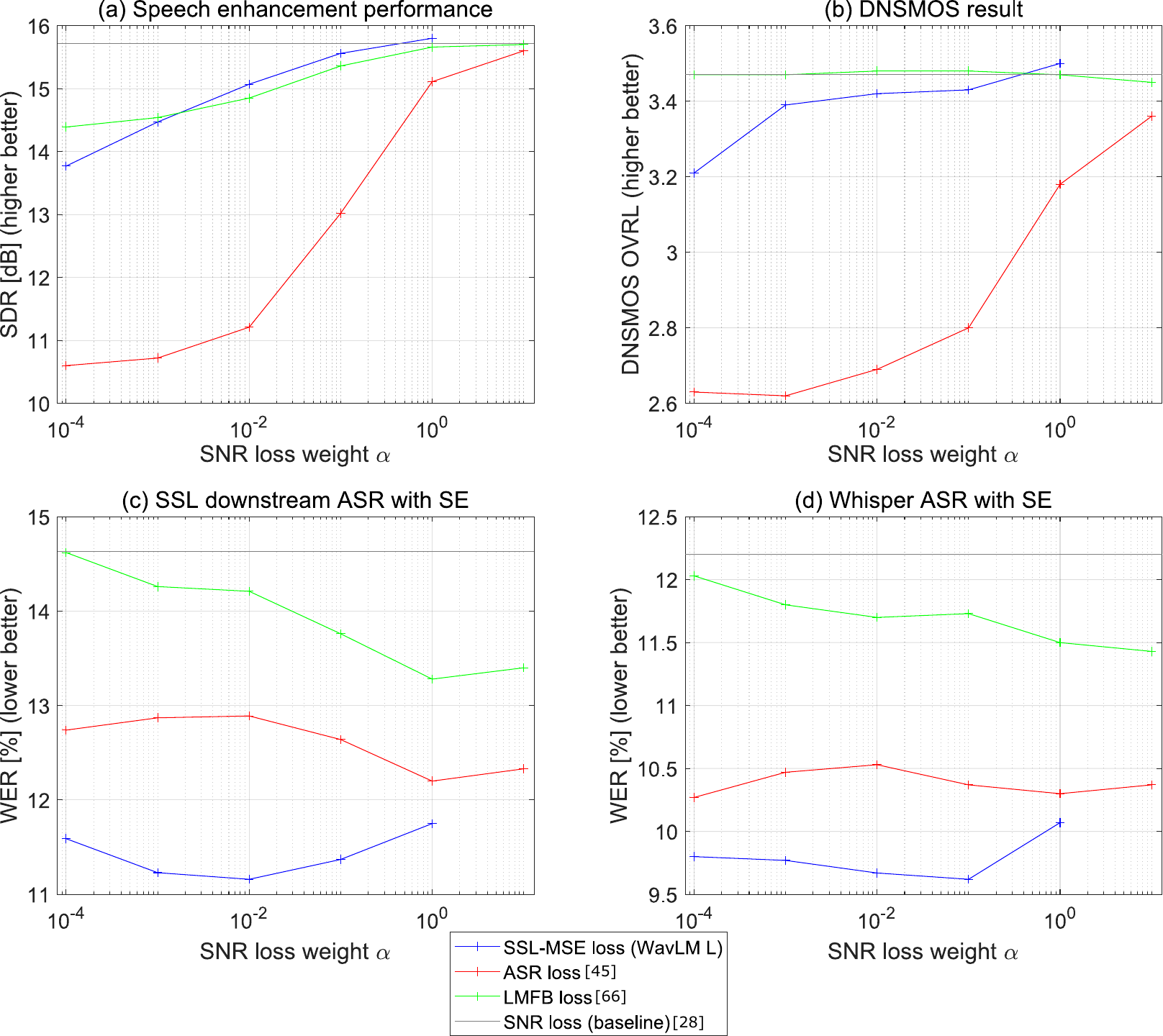}
 \end{center}
\caption{Comparison of SSL-MSE loss with ASR loss and LMFB loss joint training. The X-axis represents the multitask loss weight $\alpha$ in the joint training with each loss, and the Y-axis represents the performance in each evaluation metric. For SDR [dB] DNSMOS OVRL, higher is better, and for WER [\%], lower is better.}
 \label{fig:comparison}
\end{figure}

Figure~\ref{fig:comparison} shows the performance of SE systems trained with vanilla SNR loss, ASR loss, LMFB loss, and SSL-MSE loss. Each graph shows an evaluation metric: SE performance in SDR [dB], objective measures of perceptual quality in DNSMOS OVRL, SSL downstream ASR task performance in WER [\%], and Whisper ASR performance in WER [\%]. WavLM {\sc Base+} is used to evaluate the SSL downstream ASR performance. For each graph, the X-axis represents the multitask loss weight $\alpha$ in the multitask training, which was fixed at $\alpha=0.1$ in the Table \ref{tab:main}. The Y-axis indicates the performance. As for (a) SE performance and (b) objective measures of perceptual quality, all the multitask-trained systems share the trend that the performance approaches baseline SNR when multitask loss weight $\alpha$ is large, while the performance drops when the multitask loss weight $\alpha$ is small. The performance drop is more pronounced with ASR loss, compared with SSL-MSE loss and LMFB loss. This degradation is likely caused by the SE model becoming overly specialized for the ASR task. 

(c) SSL-based ASR performance and (d) Whisper ASR performance are both improved by incorporating SSL-MSE, ASR loss, and LMFB loss, compared with SNR loss. Among these, the SSL-MSE loss yields the most substantial improvement, followed by the ASR loss. Although the ASR loss is task-matched, it does not provide the greatest performance gains. Compared to the Whisper ASR performance without SE (9.3\%) shown in Table~\ref{tab:main} (a2), the ASR loss actually degrades Whisper ASR performance across all multitask loss weight settings $\alpha$. This degradation appears to result from the SE model becoming overly specialized to the specific ASR model used during ASR-loss training, which differs from the Whisper ASR model, as described in Section~\ref{sec:multitaskloss}. In other words, the ASR loss leads to insufficient generalization not only across different tasks, but also across different ASR models.
Therefore, SSL-MSE loss improves SE generalizability to different tasks better than other common training criteria of ASR loss and LMFB loss.

The effect of the SSL-MSE loss on (c) SSL downstream ASR and (d) Whisper ASR performance is maximized when the multitask loss weight $\alpha$ is set to 0.01. In contrast, (a) the SDR and (b) DNSMOS OVRL scores increase with larger values of $\alpha$, indicating a trade-off between human perceptual quality and downstream task performance.

\subsubsection{Generalizability to other types of SSL models}
\label{sec:cross}
\begin{table}[tpb]
\centering
\sisetup{detect-weight,mode=text}
\renewrobustcmd{\bfseries}{\fontseries{b}\selectfont}
\renewrobustcmd{\boldmath}{}
\newrobustcmd{\BOLD}{\bfseries}
\caption{Performance variations due to the combination of SSL model for inference and SSL model for training with SSL-MSE loss. w2v2 B, Hubert B, WavLM B+, and WavLM L stands for wav2vec 2.0 Base, Hubert Base, WavLM Base+, and WavLM Large, respectively. The SNR multitask loss weight $\alpha$ for SSL-MSE loss training is fixed to 0.1.}
\label{tab:cross}
\begin{tabular}{@{}lccc|S[table-format=2.1]S[table-format=2.1]S[table-format=2.1]@{}}\toprule
 & \multirow{2}{*}{\begin{tabular}[c]{@{}c@{}}SSL model\\ for inference\end{tabular}} & \multirow{2}{*}{SE} & \multirow{2}{*}{\begin{tabular}[c]{@{}c@{}}SSL model\\ for SSL-MSE \\ loss\end{tabular}} & ASR & ASV & IC \\
 & & & & {\begin{tabular}[c]{@{}c@{}}WER↓\\ \text{[\%]}\end{tabular}} & {\begin{tabular}[c]{@{}c@{}}EER↓\\ \text{[\%]}\end{tabular}} & {\begin{tabular}[c]{@{}c@{}}Acc↑\\ \text{[\%]}\end{tabular}} \\ \midrule
(a2) & WavLM B+ & & & 17.1 & 10.9 & 67.4 \\
(b) & WavLM B+ & \checkmark & N/A & 14.6 & 8.6 & 84.8 \\
(d1) & WavLM B+ & \checkmark & w2v2 B & 14.4 & 8.6 & 85.4 \\
(d2) & WavLM B+ & \checkmark & Hubert B & 14.7 & 8.7 & 86.1 \\
(c1) & WavLM B+ & \checkmark & WavLM B+ & 12.9 & 8.1 & 87.7 \\
(c2) & WavLM B+ & \checkmark & WavLM L & \BOLD 11.3 & \BOLD 7.4 & \BOLD 88.6 \\ \midrule
(e1) & w2v2 B & & & 36.8 & 19.0 & 41.0 \\
(e2) & w2v2 B & \checkmark & N/A & 18.8 & 12.2 & 70.9 \\
(e3) & w2v2 B & \checkmark & WavLM B+ & 17.5 & 11.3 & 71.3 \\
(e4) & w2v2 B & \checkmark & WavLM L & \BOLD 14.9 & \BOLD 11.0 & \BOLD 75.9  \\
(e5) & w2v2 B & \checkmark & w2v2 B & 18.5 & 12.5 & 71.2 \\ \midrule
(f1) & Hubert B & & & 31.7 & 15.5 & 57.8 \\
(f2) & Hubert B & \checkmark & N/A & 18.5 & 10.6 & 81.9 \\
(f3) & Hubert B & \checkmark & WavLM B+ & 16.9 & 9.7 & 82.6 \\
(f4) & Hubert B & \checkmark & WavLM L & \BOLD 14.0 & \BOLD 8.7 & \BOLD 85.1 \\
(f5) & Hubert B & \checkmark & Hubert B & 18.8 & 11.1 & 82.3 \\ \midrule
(g1) & WavLM L & & & 9.9 & 9.8 & 52.1 \\
(g2) & WavLM L & \checkmark & N/A & 7.8 & 9.8 & 86.6 \\
(g3) & WavLM L & \checkmark & WavLM B+ & 7.2 & \BOLD 9.3 & 88.3 \\
(g4) & WavLM L & \checkmark & WavLM L & \BOLD 6.1 & 10.1 & \BOLD 89.2 \\ \bottomrule
\end{tabular}
\end{table}

Table \ref{tab:cross} shows the SSL downstream task performances for a variation of a combination of SSL models that is used to calculate SSL-MSE loss and that is used for the inference of SSL downstream tasks. 
Overall, WavLM {\sc Base+} and WavLM {\sc Large} demonstrate superior performance compared to other SSL models as the SSL models for inference. This is likely because these models are pre-trained with noise-aware strategies, making them more effective on test sets with additive noise.
For the SSL pipeline with WavLM {\sc Base+}, SSL-MSE loss calculated with WavLM {\sc Large} (c2) most improves the performance, and SSL-MSE losses calculated with wav2vec 2.0 {\sc Base} (d1) and Hubert {\sc Base} (d2) do not improve the performance compared with vanilla SNR loss training (b). Moreover, SSL-MSE losses calculated with wav2vec 2.0 {\sc Base} and Hubert {\sc Base} do not improve matched condition where the same SSL model is used for inference, which is shown in (e5) and (f5). 

On the other hand, SSL-MSE loss calculated with WavLM {\sc Base+} and WavLM {\sc Large} improves every SSL pipeline with different types of SSL models including WavLM {\sc Base+} shown in (c1) and (c2), Wav2vec 2.0 {\sc Base} shown in (e3) and (e4), Hubert {\sc Base} shown in (f3) and (f4), and WavLM {\sc Large} (g3) and (g4), compared with the SE model trained with SNR loss. 
Thus, it can be said that SSL-MSE training potentially generalizes over different types of SSL models. Compared with SSL-MSE loss calculated with WavLM {\sc Base+}, WavLM {\sc Large} offers better performance for the SSL pipeline with different SSL models. The higher generalizability of WavLM {\sc Large} compared with WavLM {\sc Base+} as the upstream model~\citep{chen2022wavlm} seems to transfer more general knowledge to the SE model via SSL-MSE loss training.

\subsubsection{The effect of SSL-MSE loss on noise-robust downstream model}
\begin{table}[tpb]
\centering
\sisetup{detect-weight,mode=text}
\renewrobustcmd{\bfseries}{\fontseries{b}\selectfont}
\renewrobustcmd{\boldmath}{}
\newrobustcmd{\BOLD}{\bfseries}
\caption{Effect of SE and SSL-MSE loss training on a noise-robust downstream model. We adopt WavLM Base plus model for SSL-MSE loss training.}
\label{tab:robust}
\begin{tabular}{@{}lcc|S[table-format=2.1]S[table-format=2.1]S[table-format=2.1]@{}}\toprule
 & \multirow{2}{*}{\begin{tabular}[c]{@{}c@{}}SE\\ method \end{tabular}} & \multirow{2}{*}{\begin{tabular}[c]{@{}c@{}}is downstream \\ model \\ noise robust \end{tabular}} & ASR & ASV & IC \\
 & & & {\begin{tabular}[c]{@{}c@{}}WER↓\\ \text{[\%]}\end{tabular}} & {\begin{tabular}[c]{@{}c@{}}EER↓\\ \text{[\%]}\end{tabular}} & {\begin{tabular}[c]{@{}c@{}}Acc↑\\ \text{[\%]}\end{tabular}} \\ \midrule
(a2) & no process & & 17.1 & 10.9 & 67.4 \\ \midrule
(h1) & no process & \checkmark & 13.6 & 9.2 & 79.4 \\
(h2) & SNR loss & \checkmark & 12.5 & 7.9 & 85.8 \\
(h3) & SSL-MSE loss & \checkmark & \BOLD 11.3 & \BOLD 7.7 & \BOLD 88.3 \\ \bottomrule
\end{tabular}
\end{table}

To further discuss the effectiveness of SE and SSL-MSE, we prepare an SSL pipeline where the downstream model is also trained to be robust to noise by the noisy paired data for each task. Table~\ref{tab:robust} shows the SSL pipeline performance for three downstream tasks. Since WavLM {\sc Base+} is used for the SSL pipeline, SSL upstream itself is also robust to noise. The table shows that the noise-robust training (h1) improved the performance of each task compared to the system without noise-robust training (a2). This is notable that the introduction of baseline SE front-end (h2) further improves these performances despite the fact that both SSL upstream and downstream are trained to be robust to noise. The introduction of the proposed SSL-MSE loss (h3) substantially improves the performance of system (h2). This indicates that SSL-MSE loss is effective for the back-end tasks that are trained to be robust to noise.

\section{Conclusion}
To build a general speech enhancement (SE) model that can be applied to various downstream tasks, we proposed a self-supervised learning mean squared error (SSL-MSE) loss training scheme that exploits the rich speech representation captured by the SSL models to train SE models. By calculating the loss term on the feature domain of the SSL model, SE is expected to be trained to preserve high-level information contained in speech audio that is crucial to solving various downstream tasks. The experimental validation shows that introducing SSL-MSE loss improves SSL downstream task performance and Whisper ASR performance while maintaining DNSMOS results. In addition to this, the combination of OA with SSL-MSE further improves the performance of back-end tasks. Further analyses show that SSL-MSE loss can generalize over other types of SSL models than that used to calculate SSL-MSE loss. It is also shown that SSL-MSE loss training requires the use of a noise-robust SSL model like WavLM {\sc Base+} to improve performance. Moreover, SSL-MSE is effective in cases where both the upstream SSL model and downstream task model are trained to be robust to noise.

This study identifies several directions for future research. First, it is important to validate the effectiveness of the proposed approach across a wider range of downstream tasks. In this work, we demonstrated the generalizability of the model trained with SSL-MSE by evaluating its performance on three types of downstream tasks: human listening tests, tasks using SSL models, and Whisper ASR. As a future direction, we plan to explore its applicability to additional tasks such as off-the-shelf voice activity detection (VAD)~\citep{webrtcvad}, speaker diarization models~\citep{bredin2023pyannote}, etc.
Second, the evaluation of objective measures of perceptual quality using actual subjective listening tests remains an essential future task. While this study employed DNSMOS to assess the perceptual quality and demonstrated that the proposed method can improve both perceptual quality and machine understanding tasks, conducting subjective listening tests with human listeners will enable a more reliable assessment.
Third, it is necessary to investigate the applicability of the proposed approach to a broader set of speech enhancement models with different characteristics. In this work, we performed experimental validation using Conv-TasNet, a widely adopted model in the field. Future work will include exploring the generalizability of the proposed approach to various real-time speech enhancement methods~\citep{hu20g_interspeech} and larger-scale speech enhancement models~\citep{wang2023tf}.

\bibliographystyle{Frontiers-Harvard} 
\bibliography{mybib}



\end{document}